\documentclass[12pt]{article}

\usepackage{caption}
\usepackage{times}
\usepackage{afterpage}
\usepackage{floatpag}
\usepackage{graphicx} 
\usepackage{color}    
\usepackage{verbatim}
\usepackage{amsmath}
\usepackage{siunitx}
\usepackage{ulem} 
\usepackage[version=4]{mhchem}
\usepackage[left]{lineno}

\newenvironment{sciabstract}{%
\begin{quote} \bf}
{\end{quote}}

\title{Bright 25-attosecond light pulses reach the one atomic unit of time}

\author
{Jingsong Gao$^{1,\ast}$, Mahmudul Hasan$^{1,\ast}$, Hao Liang$^{2,\ast}$, \\Ming-Shian Tsai$^{3}$, Yiming Yuan$^{1}$, Zach Eisenhutt$^{1}$, Christoph H. Keitel$^{2}$, \\Chii-Dong Lin$^{1}$, Yunquan Liu$^{4,\dagger}$, Ming-Chang Chen$^{3,\dagger}$, Meng Han$^{1,\dagger}$
\\
\normalsize{$^{1}$J. R. Macdonald Laboratory, Kansas State University, Manhattan, Kansas, 66506, USA}\\
\normalsize{$^{2}$Max-Planck-Institut für Kernphysik, Heidelberg, 69117, Germany}\\
\normalsize{$^{3}$Institute of Photonics Technologies, National Tsing Hua University, Hsinchu 300044, Taiwan}\\
\normalsize{$^{4}$State Key Laboratory for Mesoscopic Physics and Frontier Science Center}\\
\normalsize{for Nano-Optoelectronic, School of Physics, Peking University, Beijing 100871, China}\\
\normalsize{$^\ast$These three authors contributed equally}\\
\normalsize{$^\dagger$E-mails: yunquan.liu@pku.edu.cn; mingchang@mx.nthu.edu.tw; meng9@ksu.edu}\\
}

\begin{document} 

\baselineskip20pt
\maketitle 
\begin{sciabstract}
Generating ever-shorter and brighter light pulses has long been a central pursuit in ultrafast science, as it benchmarks our ability to create and manipulate the coherence on the intrinsic timescale of sub‑atomic electron motion. The current state-of-the-art in attosecond pulse generation reaches durations of 40–50 attoseconds (1 as = $10^{-18}$ seconds), produced via high-order harmonic generation (HHG) driven by secondary mid-infrared light sources. However, these sources often suffer from low stability and poor HHG conversion efficiency. In this work, we demonstrate the generation of 25$\pm$2 attosecond light pulses—a new world record for the shortest light pulse—driven by a post-compressed, industrial-grade Yb-based laser system. The resulting high-harmonic spectrum spans photon energies from 50 eV to 320 eV, covering the carbon K-edge, with a calibrated photon flux exceeding $10^{12}$ photons per second, approximately three orders of magnitude higher than previous studies. The pulse duration was characterized using an angle-resolved photoelectron streaking camera on helium atoms and systematically optimized through the use of dielectric filters of varying thicknesses to compensate the attochirp. Our study reaches the threshold of one atomic unit of time (24.2 attoseconds)—the boundary between atomic and ionic physics—opening the door to resolving exciting ionic quantum dynamics with tabletop lasers.
\end{sciabstract}


The advent of attosecond light pulses \cite{paul2001observation,hentschel2001attosecond} and the parallel development of attosecond metrology \cite{itatani2002attosecond,goulielmakis2004direct} have opened the door to observing and controlling sub‑atomic motion. The attosecond timescale coincides naturally with electron dynamics in atoms: one atomic unit of time equals 24.2 as—the time it takes a hydrogen 1s electron to traverse a Bohr radius. Thanks to increasingly versatile attosecond tools \cite{nisoli1997compression,baltuvska2003attosecond,popmintchev2012bright}, researchers now probe ultrafast electron dynamics not only in gases \cite{schultze2010delay,isinger2017photoionization,villeneuve2017coherent,grundmann2020zeptosecond} but also in liquids \cite{jordan2020attosecond,luu2018extreme} and solids \cite{cavalieri2007attosecond,tao2016direct,ndabashimiye2016solid}, extending the reach of ultrafast science far beyond traditional atomic, molecular and optical physics. Resolving even faster electronic motion now hinges on producing ever shorter, more intense isolated attosecond pulses, driving a growing demand for brighter light sources at the atomic‑unit timescale. 

In Fig. 1A we summarize previous efforts on table-top attosecond light pulse generation and characterization, focusing specifically on pulse duration. For a comprehensive review, see Ref. \cite{midorikawa2022progress}. In the first ten years, attosecond sources relied exclusively on Ti:Sa lasers \cite{hentschel2001attosecond,kienberger2004atomic,sansone2006isolated,takahashi2008coherent,goulielmakis2008single,wirth2011synthesized,zhao2012tailoring}. These systems typically produced 20–50 fs pulses at 800 nm that were subsequently compressed to 4–5 fs by gas‑filled hollow‑core fibers \cite{nisoli1997compression} and chirped multilayer mirrors. The photon energy of the phase-matched HHG driven by these lasers are limited to 150 eV and the pulse duration is limited to about 60 as \cite{zhao2012tailoring}. Achieving shorter pulses requires a broader spectral bandwidth. In HHG, the single-atom cutoff energy scales with the product of the driving intensity ($I$) and the square of the wavelength ($\lambda$), i.e., $I\lambda^2$. Consequently, the second decade of attosecond light source development has been driven primarily by secondary mid-infrared pulses, produced via optical parametric amplification (OPA) \cite{teichmann20160,gaumnitz2017streaking,cousin2017attosecond,rossi2020sub} or optical parametric chirped-pulse amplification (OPCPA) \cite{popmintchev2012bright,ishii2014carrier,li201753}. These sources, typically centered around 1.8 $\mu$m, can generate photons into the soft X-ray regime, including the water window (284–530 eV), enabling element-specific spectroscopic applications. However, HHG driven by mid-infrared sources suffers from low conversion efficiency, which scales unfavorably with wavelength as $\eta \sim 1/\lambda^{5\sim7}$ \cite{midorikawa2022progress}, primarily due to the increased spatial spread of the freed electron wavepacket. Recently, industrial-grade Yb lasers \cite{balvciunas2011carrier,jeong2018direct,muller2021multipass,hadrich2022carrier,jeong2022guiding,beetar2020multioctave,tsai2022nonlinear,chien2024filamentation,pi2025synthesis} centered at 1030 nm attracted much attention due to its high repetition rate, average power, and robustness (see Ref. \cite{truong2025few} for a recent review). A common assumption is that the 1030-nm drivers cannot effectively reach the water-window spectral range under typical gas pressure conditions. This is because simply increasing the laser intensity does not continuously extend the HHG cutoff, as over-ionization disrupts phase matching. We overcome this limitation by using near-single-cycle driving pulses. Figure 1B shows the calculated phase-matched cutoff energies as a function of pulse duration for helium (He), neon (Ne), and argon (Ar) targets. The results predict that phase-matched harmonics from helium can exceed 300 eV when the driving pulse is compressed below 5 fs—a regime accessible via our cascaded post-compression technique \cite{tsai2022nonlinear}. In this work, we demonstrate the generation of isolated soft-X-ray (SXR) attosecond pulses spanning 50–320 eV, driven by carrier-envelope-phase (CEP)-stabilized 3.7-fs, 0.8-mJ pulses at a 10 kHz repetition rate. The calibrated flux exceeds $10^{12}$ photons/s at central energies around 150 eV, about three orders of magnitude brighter than previous SXR sources \cite{li201753}. 

Characterizing broadband SXR attosecond pulses presents two major challenges. The first is the spectral overlap of photoelectrons originating from the ionization of multiple atomic orbitals in the streaking characterization experiments. The only viable solution is to use a helium target, which possesses a single atomic orbital. However, helium suffers from an extremely low photoionization cross section in the SXR regime. Combined with the inherently low photon flux of SXR attosecond pulses, this makes accurate photoelectron streaking measurements with helium particularly difficult. Figure 1C compares the photoionization cross sections of helium and neon, a commonly used target in the streaking. At 150 eV, helium’s cross section is more than an order of magnitude lower than those of neon \cite{cousin2017attosecond}. Prior to this work, the 43-attosecond pulse was characterized using a xenon target \cite{gaumnitz2017streaking}, which required modeling the contribution from multiple orbitals during pulse retrieval. The second challenge lies in the use of angle-averaged or partially angle-integrated detection systems, such as widely used time-of-flight (TOF) spectrometers. Unlike RABBITT measurements \cite{paul2001observation}, attosecond streaking traces oscillate in opposite energy directions across the two momentum hemispheres. Partial angle integration thus introduces cancellation effects and complicates the pulse reconstruction. In this study, we overcame both challenges by performing angle-resolved photoelectron streaking on helium using a thick-lens, high-energy velocity-map imaging (VMI) spectrometer, combined with our brightest tabletop SXR source and stable in-line attosecond beamline. We systematically recorded photoelectron streaking traces using variable-thickness filters to compensate for the intrinsic attochirp. The shortest pulse we characterized is 25 $\pm$ 2 as, establishing a new world record and, for the first time, approaching the fundamental one atomic unit (24.2 as) barrier of time.

The Volkov Transform Generalized Projection Algorithm (VTGPA) \cite{keathley2016volkov} was developed to characterize broadband attosecond light pulses without relying on the central-momentum approximation—a simplification that converts the streaking problem into a general, fast-solvable FROG problem. While the feasibility of the VTGPA has been demonstrated in several studies \cite{keathley2016volkov, gaumnitz2017streaking, gaumnitz2018complete}, its computational efficiency remains a concern. In this work, we enhance the method by replacing the generalized projection algorithm with a quasi-Newton optimization scheme. The resulting approach, which we term the Volkov Transform Quasi-Newton Algorithm (VTQNA), significantly accelerates convergence, reducing the reconstruction time from several days to less than 20 minutes on a standard desktop computer. This improvement represents a key step to characterize broadband attosecond light pulses at the atomic-unit timescale.

\section*{In-line attosecond streaking beamline}
We begin with industrial-grade CEP-stabilized 170-fs 2-mJ Yb laser pulses (Pharos, Light conversion) and compressed the pulses down to 3.7 fs with cascaded gas cells \cite{tsai2022nonlinear}. See Refs. \cite{han2025hearing,liang2025waveform} for laser details. Our stable in-line attosecond beamline is illustrated in Fig. 1D. The laser beam (7 mm in diameter) is first directed through a 1-mm-thick fused silica plate with a 2-mm central hole, which serves as a perforated beam splitter. The outer annular portion of the beam (used to drive HHG) is temporally delayed relative to the inner portion (used as the streaking or “dressing” field). Both beams are then reflected by a concentric, two-component flat mirror set: the 2-mm-diameter inner mirror is mounted on a closed-loop delay stage and reflects only the central beam. The two beams are focused into a semi-infinite gas cell filled with 2.5 bar of helium using a spherical silver mirror with a focal length of 25 cm. Notably, the semi-infinite gas cell is crucial for bright HHG in helium, as the filamentation effect enables self-guided propagation and extends the interaction length with the gas \cite{chien2024filamentation}. After the HHG generation, we used a 1.0-mm pinhole to block the residual pulse from the driving IR beam. Approximately 1 meter downstream, a variable-thickness metal or carbon filter (2.0 mm in diameter) is mounted on a perforated fused silica plate identical to the beam splitter. The inner dressing IR beam, due to its larger divergence angle compared with the SXR beam, passes through the outer region of the fused silica, thereby compensating for the temporal delay relative to the SXR beam introduced by the perforated beam-splitter plate. Figure 1E highlights the three key optical elements that enable precise in-line synchronization and beam recombination. The combined SXR attosecond pulse and IR dressing field are then focused onto a helium gas target inside a high-energy velocity-map imaging (VMI) spectrometer \cite{kling2014thick}, using a Ni-coated toroidal mirror with a focal length of 500 mm. A custom-built flat-field spectrometer equipped with a 2400 lines/mm grating (SHIMADZU 30-003), along with a 10-cm-long microchannel plate (MCP) and phosphor screen, is used to measure the full SXR spectrum without requiring any mechanical movement. For photon flux calibration, a photodiode (not shown) is placed in the spectrometer chamber before the slit. The calibrated flux exceeds $10^{12}$ photons/s, which is significantly higher than the previously reported value of $5*10^{9}$ photons/s for the 53-attosecond pulse \cite{li201753}.

\begin{figure}[htbp]
\centering
\includegraphics[width=16.5cm]{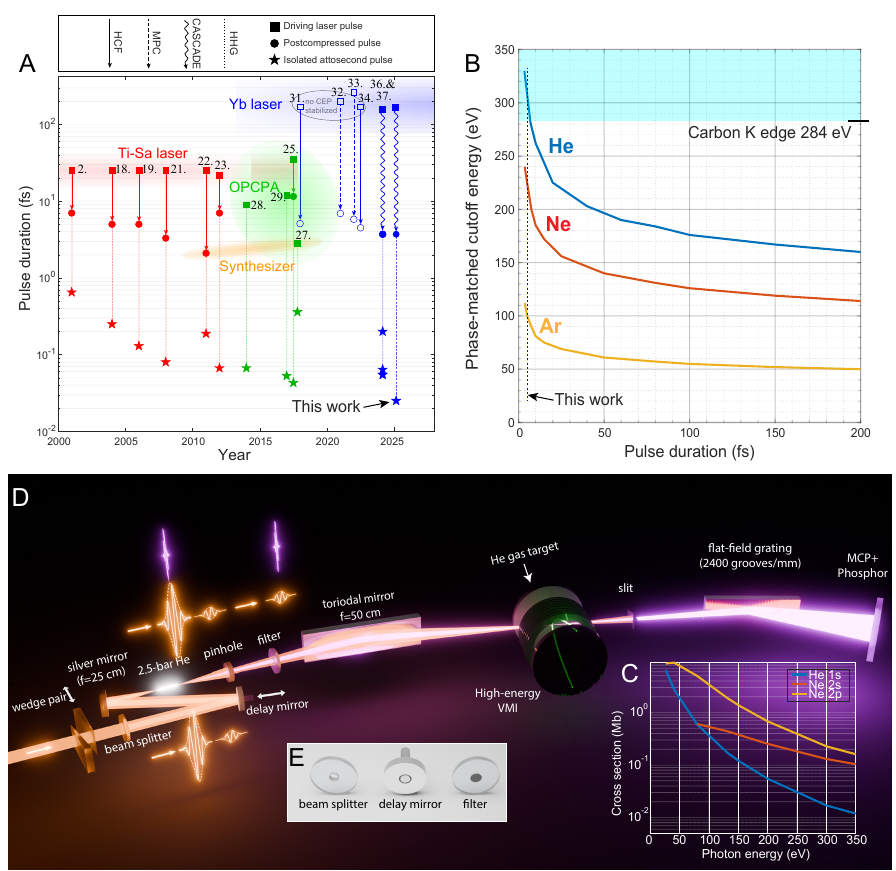}
\caption{\textbf{Atomic-unit-level attosecond light pulse generation.} (\textbf{A}) Progress in pulse duration for table-top isolated attosecond light sources. Different postcompression techniques, hollow-core fiber (HCF), multi-pass cell (MPC) and cascaded gas cells (CASCADE) are distinguished with different lineshapes. (\textbf{B}) Phase-matched cut-off energies for Ar, Ne, and He targets in HHG driven by Yb lasers centered at 1030 nm, plotted as a function of driving pulse duration. See SM for calculation details. (\textbf{C}) The comparison of photoionization cross sections between He 1s orbital and Ne 2s and 2p orbitals. Data are from Ref. \cite{yeh1985atomic}. (\textbf{D}) Schematic of our in-line attosecond streaking beamline. (\textbf{E}) The three key components are highlighted. Both the beam splitter and the filter substrate are 1-mm-thick fused silica with a central hole of 2 mm in diameter. The diameter of the inner mirror in the delay mirror set is also 2 mm.}
\label{fig:figure1}
\end{figure}

\section*{Attosecond SXR pulse reaching the carbon K edge}
Figures 2A-D illustrate our CEP-resolved spectrum of our SXR pulses measured with four different filters: 200-nm-thick tin (Sn), zirconium (Zr), aluminum (Al), and carbon (C). The corresponding transmission curve for each filter is overlaid in its respective panel. The Sn filter, which transmits photon energies above 120 eV, was previously used in the streaking characterization of the 53-attosecond pulse \cite{li201753}. In Figure 2A, the signal at 70 eV originates from the second-order diffraction of the spectral peak at 140 eV by the SXR grating. The Zr filter, shown in Figure 2B, transmits from 60 eV and exhibits relatively flat transmission around 150 eV. It was used in the characterization of the 43-attosecond pulse \cite{gaumnitz2017streaking}. Notably, the rising and falling edges of a filter's transmission curve typically introduce opposite group delay dispersion (GDD). As a result, the Zr filter is suitable only for compensating attosecond pulses below 150 eV. The Al filter, shown in Figure 2C, is commonly used either below the Al L-edge (73 eV) or above 200 eV, making it incompatible with our current spectral range. Among the filters tested, the carbon filter provides the broadest usable bandwidth in our case, transmitting from 50 eV up to the carbon K edge at 284 eV (Figure 2D). It thus enables full access to the almost entire spectral range of our generated SXR pulses. Note that our HHG spectrometer resides in an old vacuum chamber, which has some residual carbon contamination. To confirm the absorption position of the carbon K edge, Figure 2E compares HHG spectra recorded with a 200-nm Sn filter alone and with a combined 200-nm Sn and 200-nm C filter. This comparison, shown on a linear intensity scale, clearly reveals the absorption edge of carbon. Figure 2F shows the Fourier-transform-limited (FTL) pulse profile obtained from the CEP-optimized HHG spectrum using the carbon filter at the relative CEP of $\pi$. The spectrum supports an FTL pulse with a full width at half maximum (FWHM) of 11.4 attoseconds, assuming zero chirp. For a Gaussian-envelope pulse, one can estimate the FTL pulse duration $\tau$ from the spectral bandwidth $\delta\omega$ with the simple formula: $\tau$ [fs] $\cdot$ $\delta\omega$[eV] = 1.8 [fs$\cdot$eV]. In the Supplementary Materials (SM), we present CEP-resolved HHG supercontinuum spectra measured from argon and neon targets. The observed cutoff energies are $\sim$100 eV for argon and $\sim$220 eV for neon, both in excellent agreement with the theoretical phase-matching predictions shown in Figure 1B. The details of the theoretical modeling are also provided in the SM. Note that compared to previous attempts \cite{tsai2022nonlinear,chien2024filamentation} using few-cycle post-compressed pulses centered at 955 nm, we optimized the central wavelength to $\sim 1 \mu$m and increased the pulse energy to extend the cutoff.

\begin{figure}[htbp]
\centering
\includegraphics[width=16cm]{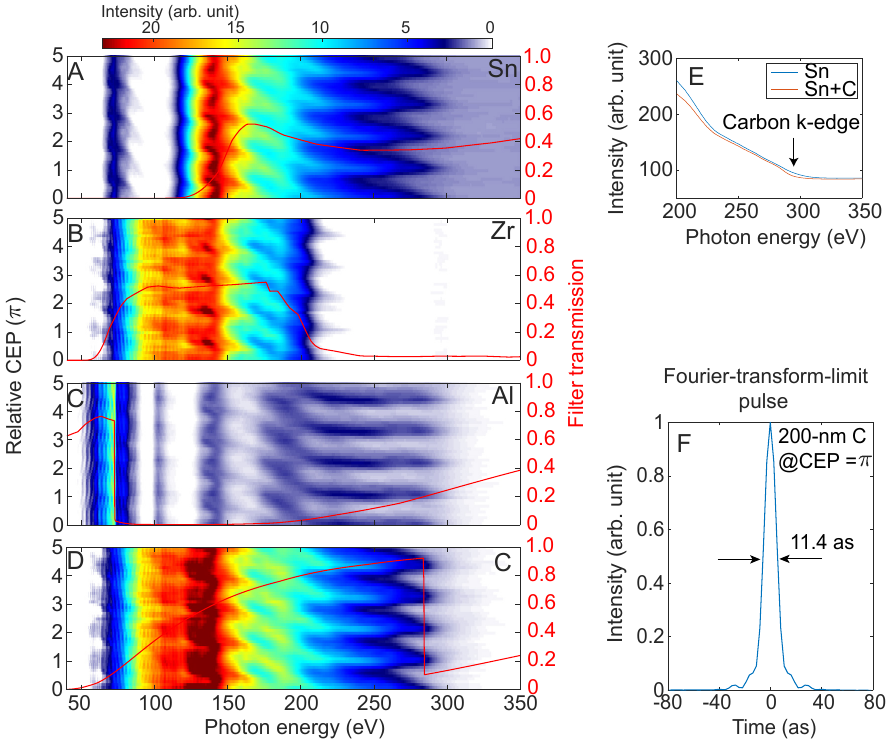}
\caption{\textbf{CEP-resolved HHG supercontinuum spectra from helium target at 2.5 bar.} (\textbf{A-D}) Measured HHG supercontinuum spectra with 200-nm-thick Sn, Zr, Al and Carbon, respectively. The solid red curves in each panels represent the transmission of the corresponding filter. The four data sets are measured with the same exposure time for the camera in the HHG spectrometer and thus their relative intensities can be compared with the same colorbar. (\textbf{E}) A comparison of HHG spectra between 200-nm Sn filter and the combination of 200-nm Sn and 200-nm C filters, which can highlight the absorption K edge of carbon in a linear scale. (\textbf{F}) Fourier-transform-limit pulse with the C filter and at the relative CEP of $\pi$.}
\label{fig:figure2}
\end{figure}

To characterize the chirp of such a broadband spectrum, we performed attosecond streaking experiments on helium atoms using our thick-lens, high-energy velocity-map imaging (VMI) spectrometer, which is optimized for detecting electrons with kinetic energies up to 350 eV and offers an energy resolution of $\Delta E/E$ better than 3$\%$ \cite{kling2014thick}. Figures 3A-D show the measured raw VMI images obtained using 200-nm Sn, Al, Zr and C filters, respectively. The bright central distribution is the secondary scattering electrons and is not the focus of this study. The outer, dipole-shape lobes are corresponding to the photoelectrons ionized from helium 1s orbital. The photoelectron spectra agree well with the features of the HHG spectra present in Figure 2. In Figure 3A (Sn filter), the outer ring corresponds to ionization signals beginning at 120 eV, consistent with the filter's transmission curve. In Figure 3B (Al filter), a sharp spectral cutoff is observed at 73 eV (Al L edge), along with a weak outer ring indicating a small contribution from higher-energy photons above 200 eV. Comparisons with the Zr and C filter data (Figures 3C and 3D) clearly reveal the Zr absorption edge at 210 eV. These observed spectral features, along with simulations of the spectrometer’s response function, allow for precise momentum calibration of the VMI system. Figures 3E–H show differential momentum distributions at selected SXR–IR time delays of –0.333, 0.167, 0.667, and 1.167 fs, respectively, spanning approximately half an optical cycle. These distributions are defined as the difference between the delay-resolved and delay-averaged momentum spectra. The results show that the entire momentum distribution is modulated—streaked up and down—by the dressing IR field, clearly revealing the angle-resolved nature of the streaking effect. This behavior is also evident in the raw VMI images, and a supplemental video is provided to further illustrate the temporal dynamics.

\begin{figure}[htbp]
\centering
\includegraphics[width=16cm]{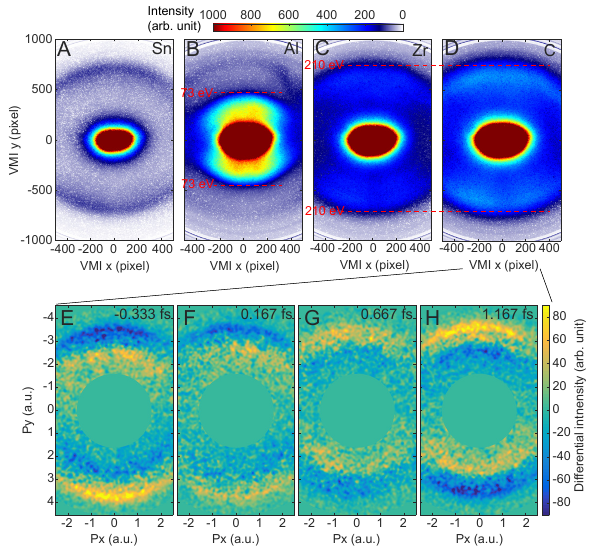}
\caption{\textbf{Angle-resolved photoelectron streaking camera}. (\textbf{A-D}) Measured raw photoelectron VMI images with 200-nm Sn, Al, Zr and C filters, respectively. In B, we mark the sharp Al absorption L edge at 73 eV. In C, we mark the Zr absorption L edge at 210 eV, which is clearly smaller than the distribution with the C filter shown in D. Note that these four data sets are also measured with the same exposure time and averaged frames, and thus their relative intensities can be compared with the same colorbar. (\textbf{E-H}) Differential momentum distributions (defined as the delay-resolved distribution minus the delay-averaged distribution) with the C filter at several SRX-IR time delays labeled on the right-top corner of each panel. We provide a supplementary video for the photoelectron streaking without subtracting the delay-averaged distribution.}
\label{fig:figure3}
\end{figure}

\section*{Compensation of SXR attochirp}

\begin{figure}[htbp]
\centering
\includegraphics[width=17cm]{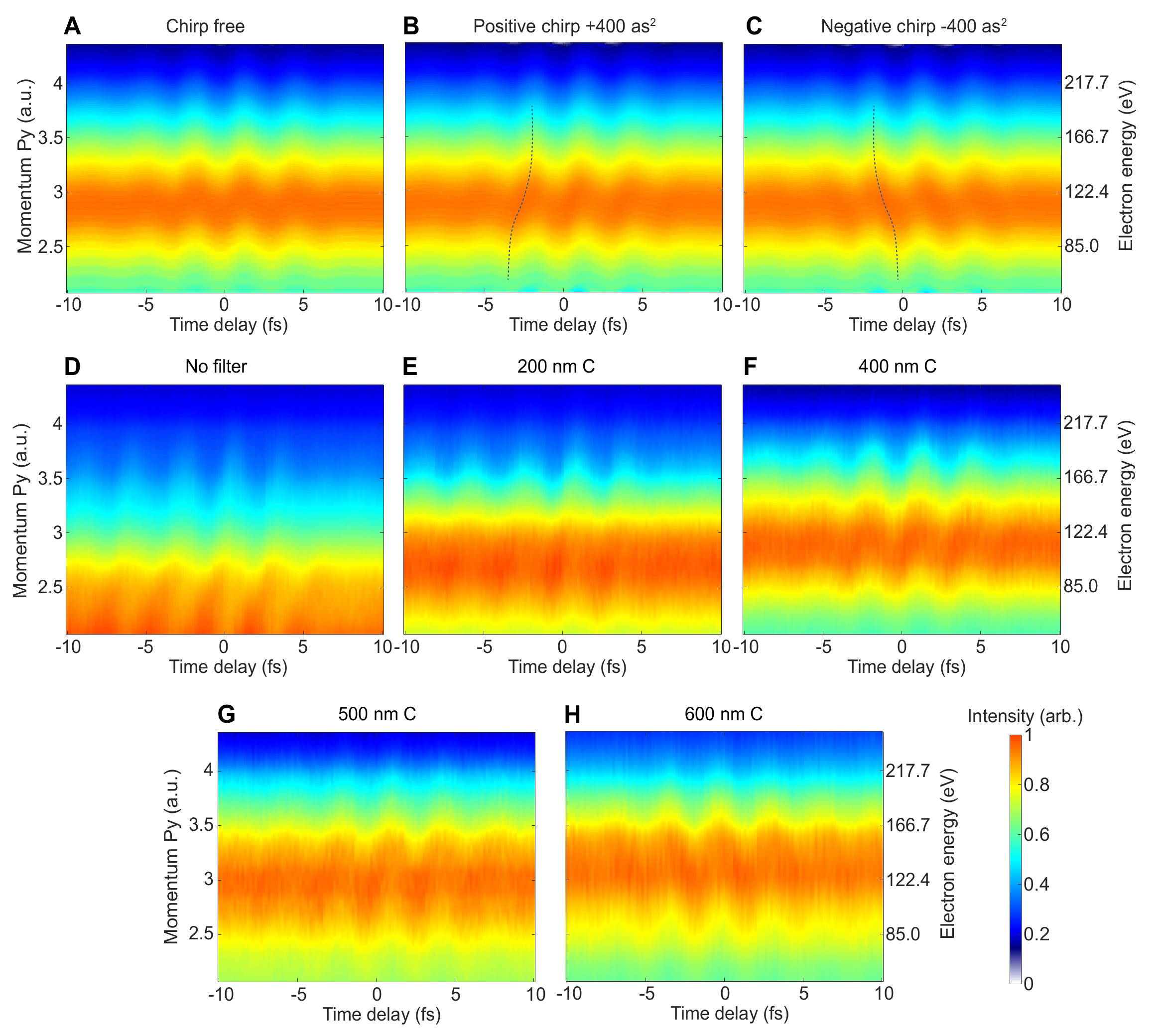}
\caption{\textbf{Attochirp compensation}. (\textbf{A-C}) Calculated streaking traces based on strong-field approximation with chirp-free, positive chirped (GDD = 400 as$^2$) and negative chirped (GDD = - 400 as$^2$) SXR pulses, respectively. A GDD of -400 as$^2$ is approximately provided by a 200-nm-thick carbon filter. (\textbf{D-H}) Measured streaking traces without any filters (D) and with 200-nm, 400-nm, 500-nm and 600-nm-thick carbon filters, respectively. The traces in all panels are normalized to their respective maxima and share the same colorbar. Note that in both calculations and measurement, Px is zero and Pz is integrated over its full range due to the VMI projection principle. Here the photoelectron streaking traces are presented in momentum-even coordinates due to the VMI detector; they would appear high-low-energy asymmetric if shown in energy-even coordinates which are typically measured by TOF spectrometers \cite{gaumnitz2017streaking,chien2024filamentation}.}
\label{fig:figure4}
\end{figure}

In the practical HHG application, phase matching is typically optimized for short electron trajectories due to their lower divergence angle and better spatial coherence. This short-trajectory phase matching inherently imparts a positive chirp to the attosecond light pulse—commonly referred to as attochirp. To achieve the shortest possible pulse duration in the time domain, this attochirp can be compensated using thin metal or dielectric filters that introduce negative GDD \cite{kim2004single,lopez2005amplitude}. To illustrate the impact of the pulse attochirp on streaking traces, Figures 4A–C present simulated results based on the strong-field approximation for three cases: chirp-free, positive chirped (GDD = 400 as$^2$) and negative chirped (GDD = - 400 as$^2$) pulses, respectively. The chirp-free SXR pulse yields a symmetric streaking trace that oscillates uniformly in phase with the vector potential of the dressing IR field. In contrast, a positively chirped pulse enhances the rising edge of the trace relative to the falling edge, while a negatively chirped pulse produces the opposite effect. These asymmetries are highlighted by the dashed curves in Figures 4B and 4C. The left–right asymmetry of the streaking trace within one optical cycle provides a direct and intuitive diagnostic for selecting the optimal filter material and thickness to compensate the attochirp.

In this study, we systematically compensate the attochirp by varying the thickness of the carbon filter. Figures 4D–H show the experimentally measured streaking traces along the same direction (upward) in the VMI spectrometer, recorded without any filter and with carbon filters of 200 nm, 400 nm, 500 nm, and 600 nm thickness, respectively. The trace recorded without any filter exhibits a more pronounced tilt with decreasing energy, indicative of a strong intrinsic attochirp, consistent with the CEP-resolved HHG spectrum shown in Figure 2D. When a 200-nm carbon filter is introduced, the streaking trace begins to resemble the oscillatory structure of the vector potential. However, the rising edge of the oscillation remains significantly more prominent than the falling edge, indicating that residual positive chirp is still present in the attosecond pulse. Increasing the filter thickness to 400 nm further reduces the chirp, yet full compensation is not achieved. In contrast, the 500-nm carbon filter yields a nearly symmetric streaking trace, suggesting that the attochirp has been effectively compensated and the pulse is nearly transform-limited. The streaking trace measured with the 600-nm carbon filter shows signs of overcompensation—manifested as a reversal in the asymmetry—indicating the introduction of a net negative chirp. As the filter thickness increases, the streaking traces gradually shift to higher energies due to the filter’s transmission characteristics. Note that, aside from the sub-cycle asymmetry observed in the streaking representation, some of current authors introduced an autocorrelation representation to intuitively assess the attochirp effect \cite{zhao2020metrology}. In the SM, we present the autocorrelation results alongside the streaking traces.

\section*{Attosecond pulse retrieval by VTQNA}

To accurately characterize the pulse duration of our broadband SXR pulses, we developed a fast-converging retrieval method: the Volkov Transform Quasi-Newton Algorithm (VTQNA). This approach replaces the generalized projection algorithm used in the original VTGPA with a quasi-Newton optimization scheme. As demonstrated in the Supplementary Materials, VTQNA reduces to VTGPA when the steepest descent iteration with an unit and fixed step size is applied. The quasi-Newton scheme improves convergence speed by at least two orders of magnitude, reducing reconstruction time to a practical level. More importantly, like VTGPA, our method does not rely on the central-momentum approximation or the assumption of single-photon transitions in the continuum. In the SM, we apply VTQNA to numerical experiments with artificial noise, demonstrating its accuracy and robustness.  

\begin{figure}[htbp]
\centering
\includegraphics[width=15cm]{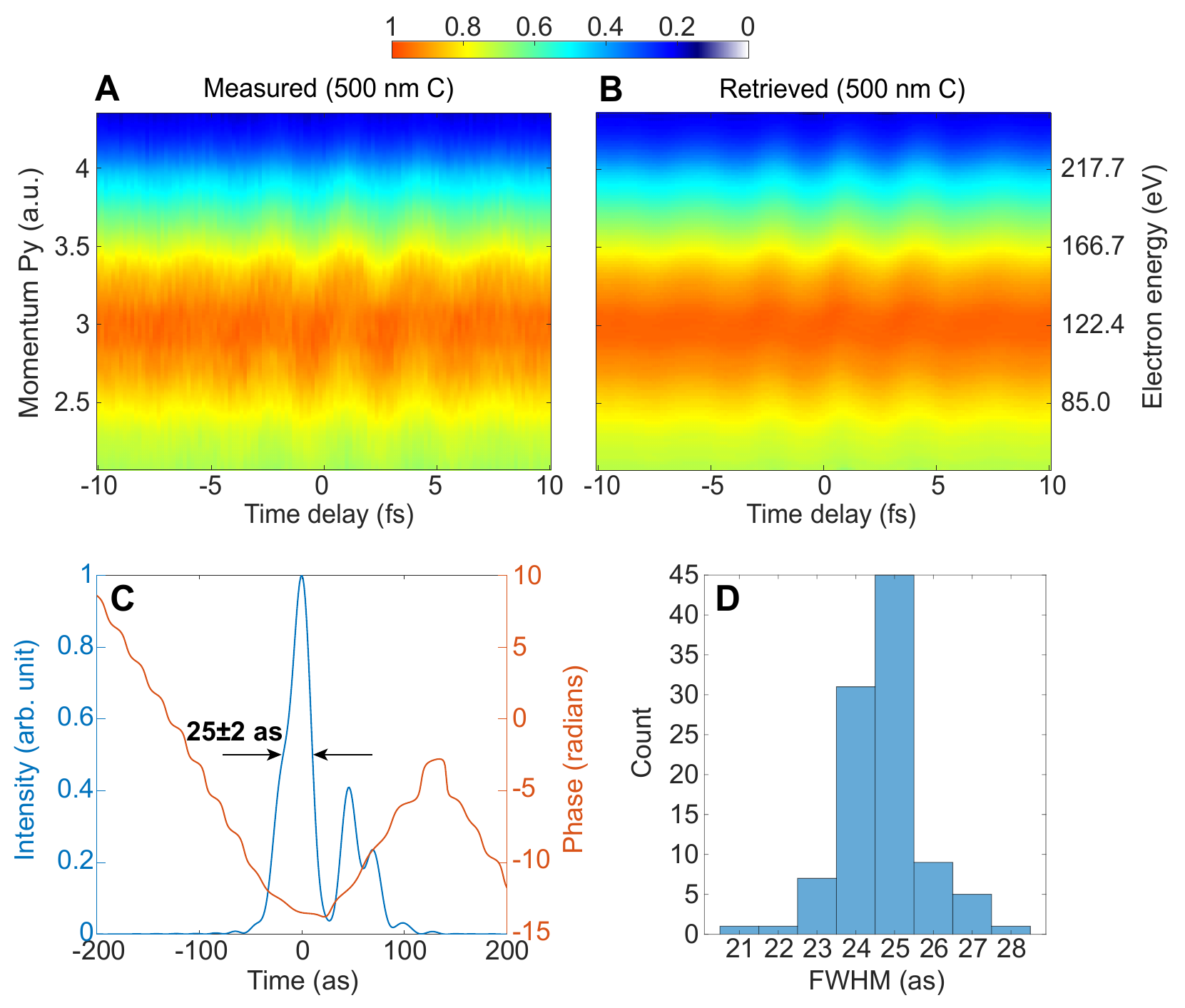}
\caption{\textbf{Attosecond pulse retrieval}. (\textbf{A-B}) Measured (A) and retrieved (B) streaking traces with a 500-nm-thick carbon filter, respectively. Note that the measured and retrieved results are presented in the same linear-scale-intensity colorbar. (\textbf{C}) Intensity envelope and temporal phase of the retrieved attosecond light pulse. (\textbf{D}) shows the histogram of the extracted full-width half maximum (FWHM). The uncertainty of $\pm 2$ as is determined by three standard deviations of the mean (i.e., 3$\sigma$).}
\label{fig:figure5}
\end{figure}

Figures 5A and 5B present a side-by-side comparison of the measured and retrieved streaking traces using the 500-nm carbon filter. The retrieved trace closely reproduces the experimental data, with an optimization merit value below $3\times 10^{-5}$, indicating excellent reconstruction fidelity. In comparison, the merit value for previous SRX pulse retrievals is typically around $4\times 10^{-3}$ \cite{gaumnitz2017streaking}. Figure 5C shows the retrieved time-domain intensity and phase of the attosecond SXR pulse obtained with the 500-nm filter. The intensity envelope reveals a dominant main pulse accompanied residual satellite pulses, which arise from high-order dispersion effects and are difficult to eliminate in broadband SXR pulses \cite{li201753}. From the intensity envelope, we extract the FWHM of the main pulse for each spectrum and present the results in the histogram shown in Figure 5D. We also retrieved the pulse duration from the streaking trace of Py $<$ 0 (see SM) and the results from both directions show good agreement. Based on this analysis, we confidently determine the pulse duration to be 25$\pm$2 attoseconds. As shown in the Supplementary Materials, the pulse retrieved using the 400-nm carbon filter exhibits a duration of 30$\pm$2 attoseconds and the result with the 600-nm carbon filter is 37$\pm$2 attoseconds, consistent with our earlier analysis of attochirp effects on the streaking traces.

In summary, we have achieved a major milestone in ultrafast science by generating attosecond light pulses that cross the threshold of one atomic unit of time. Surpassing this limit enables, in principle, the resolution of all valence electron dynamics in atomic physics, and lays the foundation for accessing ionic quantum dynamics as an essential intermediate regime en route to nuclear motion. Notably, the photon flux of our 25-attosecond pulses exceeds $10^{12}$ photons per second—a three-order-of-magnitude improvement over any previous table-top soft X-ray source. This unprecedented brightness, combined with a stable, in-line, angle-resolved attosecond streaking camera, enables new experimental capabilities, including the precise measurement of partial-wave-resolved photoionization time delays. Our results not only establish a new performance frontier but also enable widespread, tabletop access to intense attosecond pulses for investigating electronic and ionic dynamics on their intrinsic timescales.


\bibliography{pop_references}
\bibliographystyle{unsrt} 

\noindent\textbf{Acknowledgments}. We thank C. Aikens, S. Chainey and J. Millette for their technical support. We thank Hans Jakob W{\"o}rner for lending us the long toroidal mirror used in the experiments. M. Han thanks Charles Lewis Cocke for fruitful discussions on the VMI detector.

\noindent\textbf{Funding}. Kansas State group was supported by the Chemical Sciences, Geosciences and Biosciences Division, Office of Basic Energy Sciences, Office of Science, US Department of Energy, grant no. DE-FG02-86ER13491. The NTHU group was supported by the National Science and Technology Council, Taiwan, under grant no. 113-2112-M-007-042-MY3.

\noindent\textbf{Authors contributions} M. Han, M.-C. C. and Y. L. conceived the study. J. G., M. Hasan, and M. Han. performed the experiments with the support from M.-C. C., M.-S. T., Y. Y. and Z. E.. H. L. performed the SFA simulations and introduced the VTQNA pulse retrieval method. M. Han, H. L., C. H. K., C.-D. L., Y. L. and M.-C. C. analyzed and interpreted the data. M. Han wrote the paper with the input from all authors.

\noindent\textbf{Competing interests} None declared. 

\noindent\textbf{Data and materials availability} All data needed to evaluate the conclusions in the paper are present in the paper or the supplementary materials.

\section*{Supplementary Materials}
\noindent A supplementary video for the raw VMI streaking without subtracting the delay-averaged distribution.

\noindent VTQNA source code together with experimental data.

\noindent This PDF file includes:

\noindent Experimental details

\noindent Simulation and experiment of phase-matched HHG cut-off energy in helium, neon and argon

\noindent VTQNA details

\noindent Figs. S1-6

\clearpage
\setcounter{page}{1}

\end{document}